\documentclass[twocolumn,showpacs,preprintnumbers,amsmath,amssymb]{revtex4}

\usepackage{graphicx}
\usepackage{amssymb}
\usepackage{hyperref}

\begin{document}
\title{Approaching Quantum Criticality in ferromagnetic
Ce$_2$(Pd$_{1-x}$Rh$_x$)$_2$In alloys}

\author{J.G. Sereni$^{1*}$, M. Giovannini$^2$, M. Gómez Berisso$^1$, A. Saccone$^2$}

\address{$^1$ Div. Bajas Temperaturas, Centro At\'omico Bariloche (CNEA) and Conicet, 8400 S.C. Bariloche, Argentina\\
$^2$ Dipartimento di Chimica e Chimica Industriale,
Universit$\grave{a}$ di Genova, I-16146 Genova, Italy}

\date{\today}

\begin{abstract}

{Low temperature magnetic and thermal ($C_m$) properties of the
ferromagnetic (FM) alloys
Ce$_{2.15}$(Pd$_{1-x}$Rh$_x$)$_{1.95}$In$_{0.9}$ were investigated
in order to explore the possibility for tuning a quantum critical
point (QCP) by doping Pd with Rh. As expected, the magnetic
transition observed at $T=4.1$\,K in the parent alloy decreases
with increasing Rh concentration. Nevertheless it splits into two
transitions, the upper being antiferromagnetic (AF) whereas the
lower FM. The AF phase boundary extrapolates to $T_N =0$ for
$x_{cr}\approx 0.65$ whereas the first order FM transition
vanishes at $x\approx 0.3$. The QC character of the $T_N \to 0$
point arises from the divergent T dependence of the tail of
$C_m/T$ observed in the $x=0.5$ and 0.55 alloys, and the tendency
to saturation of the maximum of $C_m(T_N)/T$ as observed in
exemplary Ce compounds for $T_N \to 0$. Beyond the critical
concentration the unit cell volume deviates from the Vegard's law
in coincidence with a strong increase of the Kondo temperature.}

\vspace{0.1cm}

$^*$ E-mail-address of corresponding author:
jsereni@cab.cnea.gov.ar

\end{abstract}

 \maketitle


\section{Introduction}

The Ce$_{2\pm u}$Pd$_{2\mp y}$In$_{1-z}$ family of alloys shows an
extended range of solid solution \cite{Mauro00} with a peculiar
magnetic behavior since the Ce-rich branch ($u>0>y$) behaves
ferromagnetic (FM) whereas the Pd-rich ($y>0>u$) is
antiferromagnetic (AF). Such a difference of magnetic structure
under a small variation of the alloy composition indicates that
the energies of both phases are very similar. Another evidence for
the competition between magnetic structures in this type of
compounds is given by Yb$_2$Pd$_2$In$_{1-x}$Sn$_x$ \cite{Bauer},
which orders magnetically at intermediate $x$ values despite
stoichiometric limits are not magnetic. Consequently this family
of alloys are appropriated materials for testing the stability of
exotic order parameters.

For this work, we have profited from the FM character of the
Ce$_{2.15}$Pd$_{1.95}$In$_{0.9}$ composition to search for a
quantum critical point (QCP) \cite{TVojta,HvL} by tuning the
chemical potential through the doping of the Pd lattice with Rh,
like in the previously studied pseudobinary compound
CePd$_{1-x}$Rh$_x$ \cite{CePdRh}. For such a purpose, we have
investigated the low temperature properties through magnetic ($M$)
and specific heat ($C_P$) measurements performed on the Rh doped
Ce$_{2.15}$(Pd$_{1-x}$Rh$_x$)$_{1.95}$In$_{0.9}$.

\section{Experimental and results}

\begin{figure}[h]
\begin{center}
\includegraphics[width=19pc]{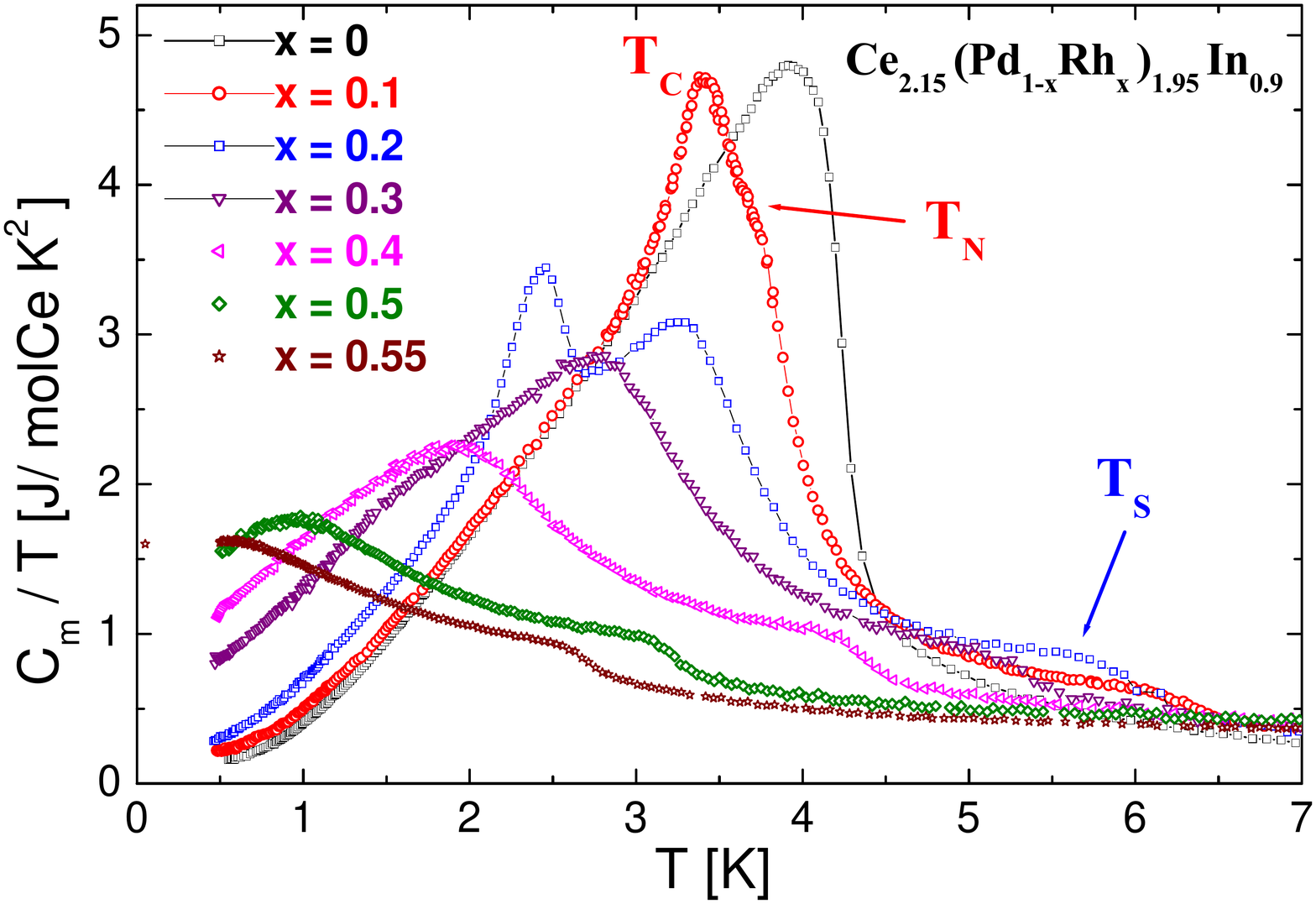}
\end{center}
\caption{Thermal and concentration dependence of the magnetic
contribution to the specific heat divided temperature. $T_C$ and
$T_N$ represent respective FM and AF transitions whereas $T_S$
identifies a concentration dependent satellite transition
discussed in the text.} \label{F1}
\end{figure}

This alloyed system forms continuously all along the complete
Pd/Rh concentration range with the Mo$_2$B$_2$Fe type crystalline
structure. The samples were prepared using a standard arc melting
procedure under an argon atmosphere, and remelted several times to
ensure good homogeneity. The range of concentration investigated
in this work covers more than $50\%$ of the total Rh by Pd
substitution, i.e. $0 \leq x \leq 0.55$. From structural
properties, one observes that the volume of the unit cell
decreases with Rh content following a Vegard's law up to $x=0.3$.
Beyond that concentration the volume decreases faster indicating
that hybridization effect become relevant. The main change
observed as a function of concentration occurs along the c-axis.

The low temperature behavior was investigated on thermal and
magnetic properties. Concerning $C_P(T)$ measurements, in
Fig.~\ref{F1} we present the thermal dependence of the magnetic
contribution divided by temperature ($C_m/T$), after phonon
subtraction for different concentrations extracted from the
non-magnetic La isotypic reference compound.

\begin{figure}[h]
\begin{center}
\includegraphics[width=18.5pc]{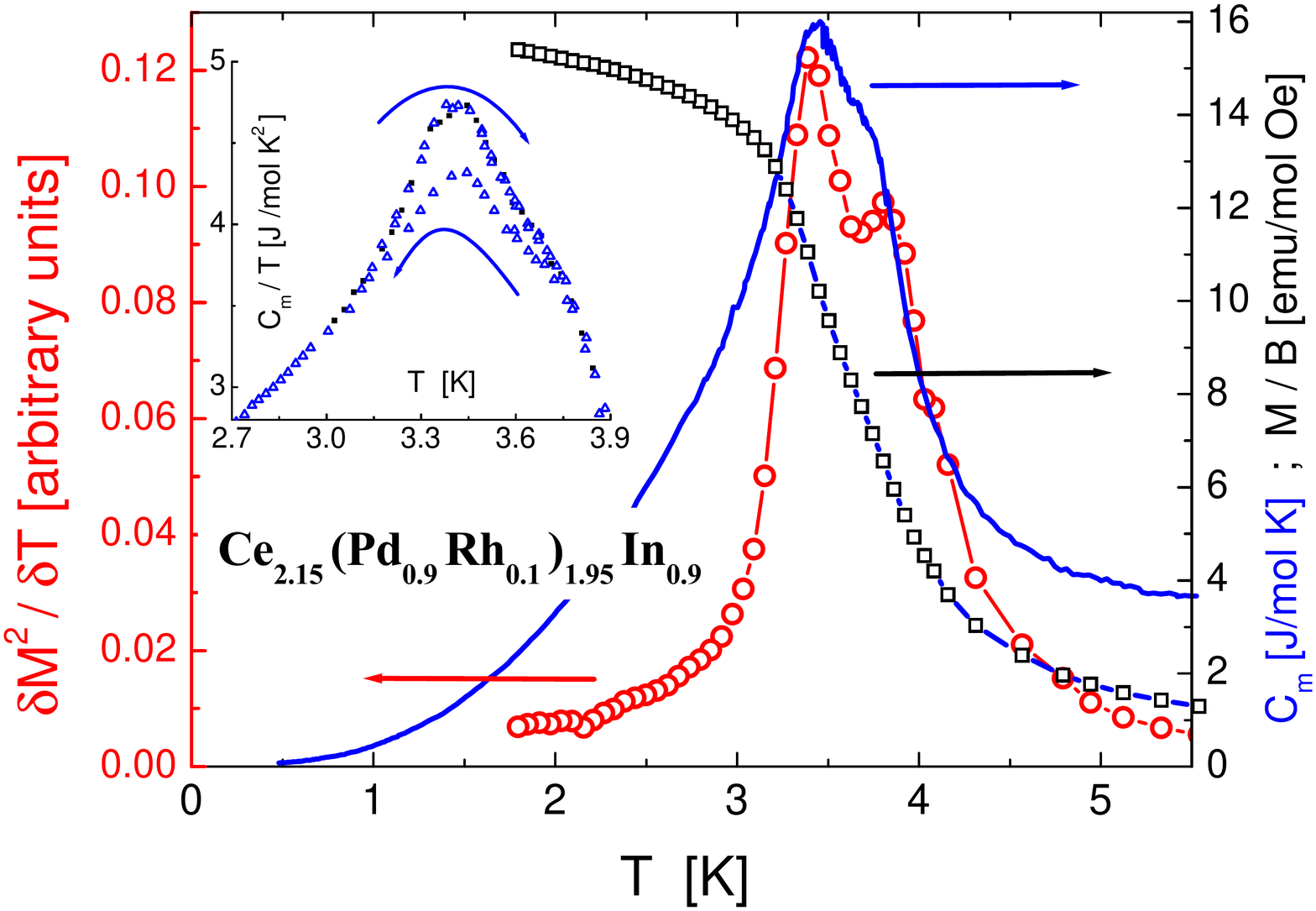}
\end{center}
\caption{Comparison between thermal derivative of $M^2(T)$ (left
axis) with $C_m(T)$ (right axis) showing the split between FM and
AF transitions. The original $M(T)$ dependence is also included
(right axis). Inset: hysteresis of $C_m/T$ between heating and
cooling procedures.} \label{F2}
\end{figure}

As it can be seen in Fig.~\ref{F1}, the magnetic transition in the
parent compound at $x=0$ shows the characteristic $C_m/T$ jump of
second order character. However, a slight broadening at the
maximum of $C_m/T$ develops, which was previously interpreted as
an intrinsic experimental broadening \cite{Mauro00}. The new
results indicate that the sample at x = 0 is actually close to a
bi-critical point at $T_{cr}=4.1$\,K since the phase boundary
splits into two transitions by Rh doping. Similar situation was
found in the well known CeRu$_2$(Ge$_{1-x}$Si$_x$)$_2$ system
\cite{Haen}. Also in Fig.~\ref{F1} a Rh concentration dependent
satellite anomaly ($T_S$) is observed. The intrinsic character of
this satellite transition is discussed below.

Magnetic $M(T)$ measurements on the $x=0$ sample do not improve
the identification of the bi-critical point because the FM signal
largely exceeds the underlying AF cusp. The presence of this
bi-critical point is not surprising if we consider a previous work
\cite{Sereni11} on the Ce$_{2\pm u}$Pd$_{2\mp y}$In$_{1-z}$ family
of alloys. From that study, one concludes that $T_{cr}=4.1$\,K,
lies in the extrapolated values of $T_N(y)$ from the Pd-rich AF
samples (i.e. $Pd\geq 2+y$) to the Ce-rich FM-branch (i.e. $Pd\geq
2-y$). The same coincidence is observed for the paramagnetic
temperature $\theta_P$ extrapolated to $Pd = 1.95$ from the
$Pd\geq 2$ side.

In the following we will analyze the evolution of the low
temperature properties for further increase of the Rh content. The
splitting between $T_N$ and $T_C$ becomes more clear in the
thermal and magnetic results from the
Ce$_{2.15}$(Pd$_{1-x}$Rh$_x$)$_{1.95}$In$_{0.9}$ sample at
$x=0.1$. That feature is observed as an incipient structure in the
cusp of $C_m/T$, see Fig.~\ref{F1} and Fig.~\ref{F2}. The first
order character of the (lower) $T_C$ transition is evidenced by an
hysteresis at the $C_m(T)$ transition which shows a shift between
heating or cooling procedures as depicted in the inset of
Fig.~\ref{F2}.

A further evidence for the splitting of both phase boundaries is
obtained from $M(T)$ measurements. Taking into account that from
thermodynamic properties the internal magnetic energy $U_m$ of a
FM phase is related to the spontaneous magnetization, i.e. $U_m
\propto M^2$, its temperature derivative $\partial M^2/\partial T$
is proportional to $C_m(T)$ \cite{Belov}. In Fig.~\ref{F2} we
compare both parameters (left axis for $\partial M^2/\partial T$
and right axis for $C_m(T)$) showing that the mentioned structure
at the transition is better defined by a $\partial M^2/\partial T$
versus T dependence. The same features are observed in the $x=0.2$
sample since the transitions increase their thermal difference (
$T_N=3.3$\,K and $T_C=2.5$\,K) as observed in Fig.~\ref{F1}.

As it can be appreciated in Fig.~\ref{F1}, the nature of both
transitions changes for $x\geq 0.3$. While $T_C(x)$ tends to
vanish becoming a weak shoulder at $x=0.3$, the $C_m(T_N)$ jump
transforms into a cusp at that concentration and then into a broad
maximum for $x\geq 0.4$ centered at 1.8\,K ($x=0.4$), 0.9\,K
($x=0.5$) and 0.5\,K ($x=0.55$) respectively. Those maxima are
followed at higher temperature by a large tail resembling the
non-Fermi-liquid behavior. Coincidentally, the value of the
$C_m/T$ maxima decrease and tends to saturate at around
1.5\,J/molK$^2$ (see Fig.~\ref{F1}) as usually observed at the
proximity of a QCP \cite{SereniHvL}. Particularly, sample $x=0.55$
follows a power law divergency $C_m/T \propto 1/(T^{1.25}+1)$ once
the $T_S$ anomaly is subtracted.

\begin{figure}[h]
\begin{center}
\includegraphics[width=18pc]{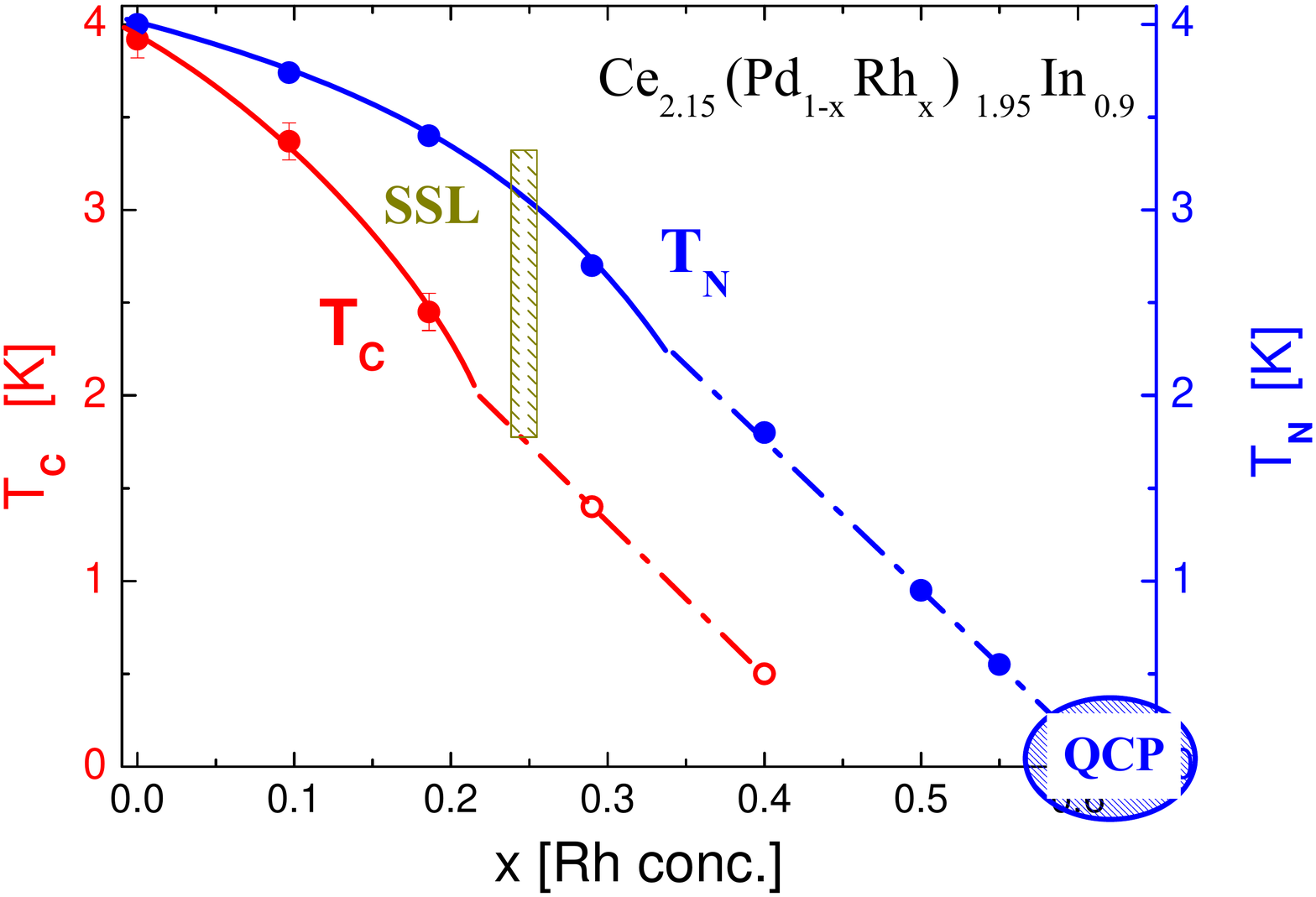}
\end{center}
\caption{Magnetic phase diagram showing $T_N(x)$ and $T_C(x)$
phase boundaries decrease. SSL indicates the region where a
Shastry Sutherland lattice is expected to form.} \label{F3}
\end{figure}

Concerning the satellite transition at $T_S(x)$, it can be mapped
out to the $T_C$ transition of the pseudo-binary compound
Ce$Pd_{1-x}$Rh$_x$ \cite{CePdRh} at a similar Rh concentration.
Coincidentally, the height of that anomaly nicely fits with the
excess of Ce respect to the stoichiometric value, c.f.
$2+u=2+0.15$, that represents a $7.5\%$ excess of Ce (remind that
this compound contains 2 Ce atoms per formula unit). This
observation is in agreement with the nearly constant entropy
contribution of the anomaly. Both, $T_S(x)$ and entropy behaviors
do not fit into a random concentration spurious contribution. This
unexpected feature can be explained by analyzing the distribution
of Ce atoms in this complex crystalline structure. As reported in
Ref. \cite{Mauro00} the Ce-plane, at $z=0.5$, contains Ce atoms
placed in the crystallographic site $4h$ whereas Ce atoms
exceeding stoichiometric concentration (i.e. $u=0.15$) replace In
atoms at the $2a$ site at the $z=0, 1$ planes. A further evidence
for the intrinsic nature of this anomaly is given by scaling
properties since for samples with $x\leq 0.4$ $C_m(x, T=T_S)$
nicely coincide if represented as a function of a normalized
temperature $t=T/T_S$. $T_S(x)$ extrapolates to $T=0$ for
$x\approx 0.75$, beyond the studied concentration range.
Preliminary magnetic measurements indicate that this anomaly is
suppressed by the application of a moderate magnetic field
($B\approx 0.1$\,T). Such a rapid suppression confirms that it is
not due to a spurious CePd$_{1-x}$Rh$_x$ contribution because that
compound has shown to behave differently under magnetic field
\cite{CePdRhOld}.

\begin{figure}[h]
\begin{center}
\includegraphics[width=18pc]{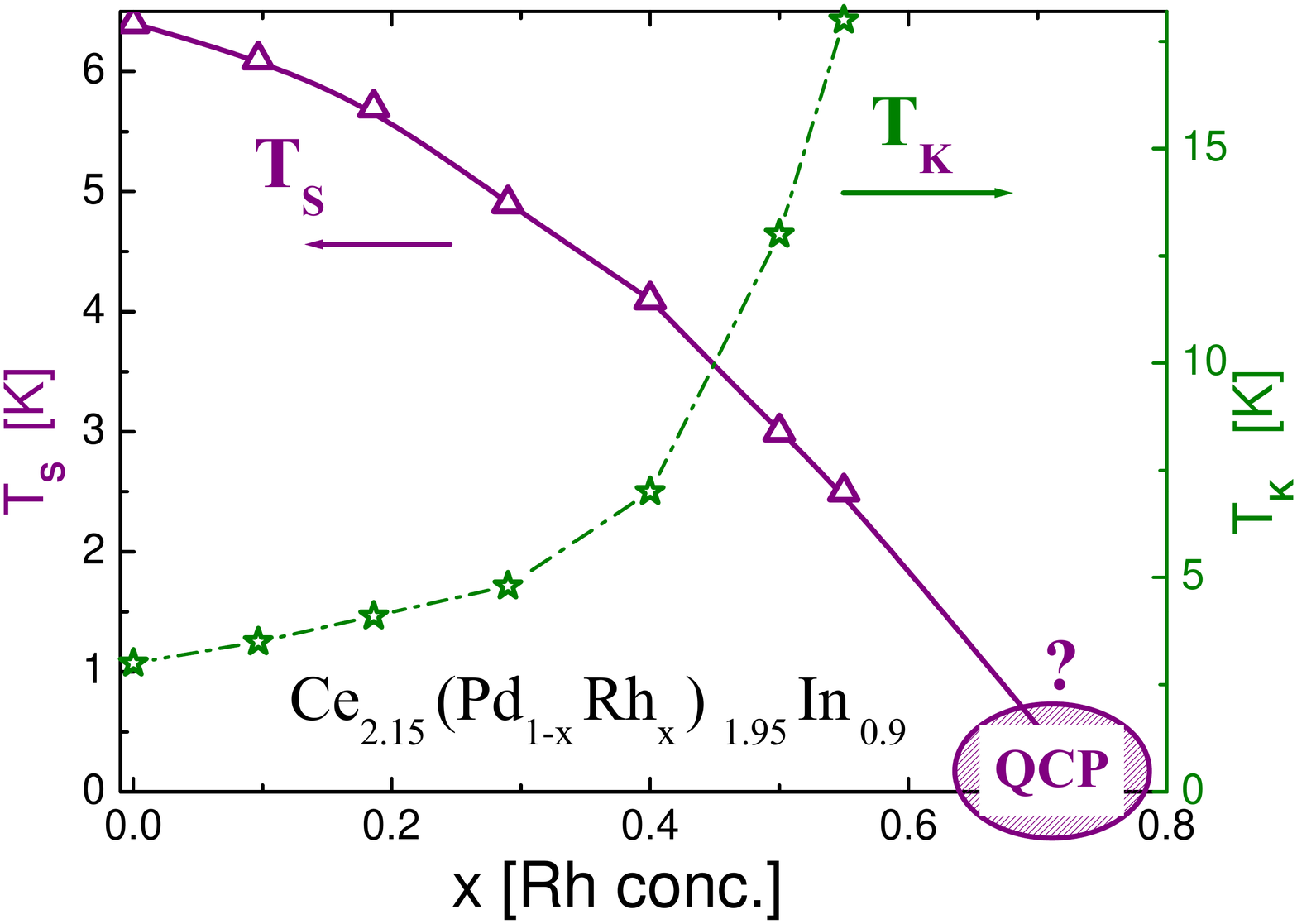}
\end{center}
\caption{Rh concentration dependence of the satellite anomaly
$T_S$ and the $T_K(x)$ increase (right axis).} \label{F4}
\end{figure}

\section{Magnetic Phase Diagrams}

In Fig.~\ref{F3} we present the phase diagram showing that the
upper ($T_N$) and the lower ($T_C$) transitions converge into a
bi-critical point at $x\to 0$. Both boundaries are well defined
for $x\leq 0.2$, but they broaden and smear respectively for
$x\geq 0.3$. Since Rh doping effect is expected to introduce {\it
holes} into the conduction band, in order to confirm the existence
of that critical point we have tested the possibility to move on
the other direction, i.e. introducing more {\it electrons} into
the band. For such a purpose we investigated the Ag doped alloy
Ce$_{2.15}$(Pd$_{0.90}$Ag$_{0.10}$)$_{1.95}$In$_{0.9}$, whose
preliminary thermal and magnetic properties indicate a clear FM
behavior with the consequent disappearance on the AF phase at the
bi-critical point.

From the magnetic behavior, the intermediate phase at $T_N \leq T
\leq T_C$ seems to behave as a Shastry-Sutherland lattice
\cite{Aronson}, as it occurs in the isotypic compound
Ce$_2$Pd$_2$Sn \cite{CeSSL}. This exotic phase, which requires of
Ce-dimers formation, is suppressed once those dimers do not form
anymore because of the weakening of Ce magnetic moments due to the
arising Kondo screening.

Fig.~\ref{F4} presents the concentration dependence of the
satellite $T_S(x)$ anomaly with its extrapolation to $x\approx
0.75$, and the Kondo temperature $T_K(x)$ evaluated using the
Desgranges-Schotte criterion \cite{Desgr}. According to this
criterion, $T_K$ can be computed as the temperature at which the
entropy reaches the value $S(T_K)=2/3RLn2$. It is worth noting
that the rapid increase of $T_K(x)$ coincides with the
extrapolation of $T_N(x)$ to the quantum critical point, and the
deviation of $V(x)$ from the Vegard's law. On the contrary, the
degrees of freedom involved in the $T_S(x)$ anomaly seems not to
be affected by Rh increase at least up to $x=0.55$, see
Fig.~\ref{F1}. Further studies at higher Rh concentration are in
progress to better determine $T_S(x)$ at lower temperature and its
dependence on magnetic field.

\end{document}